\documentclass[preprint,showpacs,superscriptaddress,preprintnumbers,amsmath,amssymb]{revtex4}

\usepackage{graphicx}
\usepackage{dcolumn}
\usepackage{bm}

\newcommand{\diff}{\mathrm{d}}

\newcommand\+{\dagger}  

\newcommand\Tr{\mathop{\mathrm{Tr}}}

\renewcommand\Im{\mathop{\mathrm{Im}}}

\newcommand\Lcal{\mathcal{L}}

\newcommand\Dcal{\mathcal{D}}

\newcommand\e{\mathrm{e}}
\newcommand\p{\partial}

\newcommand\B{{\rm B}}
\newcommand\F{{\rm F}}
\newcommand\C{{\rm c}}
\newcommand\Le{{\rm L}}
\newcommand\R{{\rm R}}

\newcommand\SU{{\rm SU}}
\newcommand\U{{\rm U}}

\newcommand {\beq}{\begin{eqnarray}}
\newcommand {\eeq}{\end{eqnarray}}

\begin{document}

\preprint{TKYNT-10-25}

\title{Topological Interactions 
of Non-Abelian Vortices \\
 with Quasi-Particles
 in High Density QCD 
}

\author{Yuji~Hirono}
 \email{hirono@nt.phys.s.u-tokyo.ac.jp}
\author{Takuya~Kanazawa}
 \email{tkanazawa@nt.phys.s.u-tokyo.ac.jp}
\affiliation{
Department of Physics, University of Tokyo, 
Hongo~7-3-1, Bunkyo-ku, Tokyo 113-0033, Japan
}
\author{Muneto~Nitta}
 \email{nitta@phys-h.keio.ac.jp}
\affiliation{
Department of Physics, and Research and Education 
Center for Natural Sciences,
Keio University, Hiyoshi~4-1-1, Yokohama, Kanagawa 223-8521, Japan
}

\date{\today}

\begin{abstract}
Non-Abelian vortices are topologically stable objects in the
 color-flavor locked (CFL) phase of dense QCD.
We derive a dual Lagrangian starting with the Ginzburg-Landau effective
 Lagrangian for the CFL phase, 
and obtain topological interactions of non-Abelian vortices
with quasiparticles such as $\U(1)_\B$ Nambu-Goldstone bosons (phonons) 
and massive gluons. 
We find that the phonons couple to the translational zero modes of the vortices 
while the gluons couple to their orientational zero modes 
in the internal space.
\end{abstract}


\maketitle

\section{Introduction}\label{sec:intro}

Topological defects are important in condensed matter physics since 
they can affect properties of matter or even the phase structure 
of the system.
This can also be the case in the condensed matter physics of QCD \cite{Rajagopal:2000wf}.
It seems likely from theoretical studies that dense and cold quark
matter is a color superconductor \cite{Alford:2007xm}.
Such a state of matter can be realized in the core of
neutron stars or possibly in heavy ion collisions.
It is quite important to investigate the properties of a color superconductor to
verify its existence experimentally and observationally.
In particular, at extremely high density and low temperature, 
perturbative calculations are reliable, 
and it is believed that color-flavor locked (CFL) phase emerges 
where all of the three light quarks make pairs and form condensates 
\cite{Alford:1998mk,Alford:2007xm}.
The original symmetry of QCD, 
$G=\U(1)_\B \times \SU(3)_\C \times \SU(3)_\Le
\times \SU(3)_\R$, is spontaneously broken down to the color-flavor locked symmetry 
$H=\SU(3)_{\rm c+L+R} \equiv \SU(3)_{\rm c+F}$ 
in the CFL phase, apart from discrete symmetry.
It exhibits superfluidity
due to the breaking of the global $\U(1)_\B$ symmetry,
as well as color superconductivity because of broken color symmetry.

By examining the topology of order parameter space $G/H \simeq \U(3)$ 
it was shown for the first time in Ref.~\cite{Balachandran:2005ev} that 
topologically stable vortices exist 
as a consequence of the symmetry breaking in the CFL phase. 
The topologically stable vortices with the lowest winding number 
in the CFL phase are non-Abelian semisuperfluid vortices.
They are called ``non-Abelian'' in the sense that 
the remaining symmetry $H$ is a non-Abelian group \footnote{
$\U(1)_{\B}$ superfluid vortices found in \cite{Forbes:2001gj,Iida:2002ev} 
have winding number one while non-Abelian semisuperfluid vortices 
have winding number 1/3, and consequently the former decay 
into three of the latter.
Color magnetic flux tubes discussed in \cite{Iida:2002ev,Iida:2004if} have 
triple amount of fluxes of non-Abelian semisuperfluid vortices, 
and are topologically and dynamically unstable.
}.
Just as a vortex in a conventional type II superconductor carries 
a magnetic flux, a non-Abelian vortex in the CFL phase carries 
a {\it color} magnetic flux. 
At the same time, this object behaves as a superfluid vortex 
as a result of superfluid properties of the color superconductor.
Therefore, when a color superconductor rotates 
faster than some critical velocity, 
these vortices are created along the axis of the rotation. 
When the rotation speed is further increased, 
the created multiple vortices should form a lattice of vortices 
as in ultra cold atomic superfluids.
As a result, 
if the density in the core of a neutron star is sufficiently high  
for the color superconducting matter to exist, 
a vortex lattice should appear 
and it should yield some physical consequences which can give a signal.
The superfluid turbulence may also occur 
if there is some mechanism which leads to an instability of ordered
structure like the Donnelly-Glaberson instability \cite{Glaberson:1974zz}
in atomic superfluids.

In order to find out what kind of collective structure of
vortices is realized, 
it is essential to determine the interaction between vortices.
This can be clarified by investigating the interaction of vortices with
quasiparticles, that mediate the force between vortices.
The interaction between vortices and quasiparticles is also useful
to study the transport properties of a color superconductor.
As quasiparticles, there appear eight gluons with masses given by the
Higgs mechanism, 
eight Nambu-Goldstone (NG) bosons (CFL mesons) associated with the chiral symmetry breaking, 
and one NG boson (phonon) due to the breaking of 
the baryon number symmetry $\U(1)_\B$.

One striking feature of non-Abelian semisuperfluid vortices is that 
they have internal degrees of freedom, which are called orientational
zero modes.
The existence of a vortex breaks the color-flavor locked symmetry
$\SU(3)_{\rm c+F}$ down to its subgroup $\SU(2) \times \U(1)$ 
around the core of the vortex. 
Consequently, there appear further NG modes confined 
inside the core of the non-Abelian vortex \cite{Nakano:2008},
which parametrize the coset space 
\begin{equation}
  \mathbb{C}P^2 \simeq \frac{\SU(3)_{\rm c+F}}{\SU(2) \times \U(1)}.
  \label{eq:CP2}
\end{equation}
There are degenerate vortex solutions with different color
magnetic fluxes, which correspond to points on the $\mathbb{C}P^2$ space.
The force between largely separated non-Abelian vortices in the CFL phase is 
independent of internal orientations \cite{Nakano:2008}. 
This can be understood by noting that the long-range force 
is mediated only by massless $\U(1)_\B$ phonons, 
which are insensitive to color fluxes.
However when the separation between vortices is relatively small, 
one expects that the force can be mediated also 
by massive particles which are not color singlet  
such as gluons, 
and consequently the short-range force can depend on 
the color fluxes or the orientational zero modes of the vortices. 

In this paper, we determine the interaction between 
a non-Abelian vortex and quasiparticles such as 
gluons and phonons.
To this end we derive a dual Lagrangian corresponding to the 
low-energy effective theory of the CFL phase by making use of a dual 
transformation.
A dual transformation is a method widely used in various fields of physics, 
which relates theories that have different Lagrangians and variables 
with the roles of equations of motion and Bianchi identities exchanged.
Under this transformation, particles and solitons typically interchange their
roles, i.e. Noether currents are interchanged with topological currents, and vice-versa.
Topological defects in the original theory are mapped to 
fundamental particles in the dually transformed description.
One of the advantages of dualization is that one can deal with the
interaction of topological defects by the method of ordinary field
theory for particles.
Dualization was first used to deal with Abelian vortices;
in the case of local vortices in the Abelian Higgs model, 
the $\U(1)$ gauge field is mapped to 
massive two-form (antisymmetric tensor) fields \cite{Sugamoto:1978ft}.
On the other hand, in the case of global (or superfluid) vortices, 
the $\U(1)$ NG mode is mapped to massless two-form fields 
\cite{Kalb:1974yc,Lee:1993ty}.
In either case, vortices are mapped to strings 
which behave as a source of two-form fields \cite{Kalb:1974yc}. 
Duality can be extended to non-Abelian gauge field theories;  
The dual transformation was applied by Seo, Okawa and Sugamoto \cite{Seo:1979id}
to $\SU(2)$ gauge theory of Higgs fields in the adjoint representation, 
which admits the so-called ${\mathbb Z}_2$ vortices.
Gauge fields are dualized to non-Abelian generalization of two-form fields 
\cite{Seo:1979id,Freedman:1980us} and the ${\mathbb Z}_2$ vortex appears 
as a source of non-Abelian two-form fields.
However the ${\mathbb Z}_2$ vortices are Abelian 
and do not have internal orientational modes, 
although the gauge fields and dualized two-form fields are non-Abelian.  

We apply the duality to non-Abelian vortices in the CFL phase, 
in which internal orientational modes reside. 
Whether or how these orientational zero modes couple to 
quasiparticles is a nontrivial question.
We show that the interaction of non-Abelian vortices with
quasiparticles naturally arises by a dual transformation.
In the dual theory, dual gluon fields are described by 
massive non-Abelian two-form fields as in \cite{Seo:1979id}.
We find that non-Abelian two-form fields are coupled to non-Abelian vortices 
through their internal orientational zero modes given in 
Eq.~(\ref{eq:CP2}).
This result is quite natural because these modes correspond one-to-one
to the color magnetic flux which the non-Abelian vortex carries. 
On the other hand, the phonons are dual to 
Abelian two-form fields coupled to non-Abelian vortices 
through their translational zero modes 
in the same way as for Abelian vortices, but the coupling strength is 
$1/3$ of that for Abelian vortices. 
Because the interaction terms do not involve the space-time metric,
they are called topological interactions. 
The dual Lagrangian obtained here provides a starting point to analyze a possible
collective structure of vortices.

This paper is organized as follows.
In Sec.~\ref{sec:LEET} we give a Ginzburg-Landau Lagrangian 
of the CFL phase.
In Sec.~\ref{sec:dual} we perform a dual transformation 
of the Ginzburg-Landau Lagrangian. 
First we take a dual of massive gluon fields to 
massive non-Abelian two-form fields. 
Then we take a dual of the $\U(1)_\B$ phonon 
to the massless two-form field.
In Sec.~\ref{sec:vor-tensor} we use a solution of 
a single non-Abelian vortex to calculate 
the vorticity tensor of the non-Abelian vortex explicitly.
Sec.~\ref{sec:summary} is devoted to summary and discussion.
In Appendix we will give some details of calculations omitted in the main text. 


\section{Low-energy effective theory of the CFL phase}\label{sec:LEET}

In this section we introduce an effective theory of the CFL phase 
and discuss the symmetry of the ground state.
We start with a time-dependent Ginzburg-Landau(GL) effective Lagrangian
\footnote{The time-dependent GL Lagrangian is valid 
when the temperature is close to the critical temperature and 
deviations from equilibrium are small.
}
for the CFL phase up to second order in time and spatial derivatives. 
The Lagrangian is written in terms of order parameters $\Phi^{\rm L}$ and
$\Phi^{\rm R}$,
\beq
  [ \Phi^{\Le(\R)}]_{a i} \sim \epsilon_{abc}\epsilon_{ijk} 
  \left<q^{\Le(\R)}_{bj} C q^{\Le(\R)}_{ck} \right>,
\eeq
where $a,b,c$ and $i,j,k$ are color and flavor indices, respectively, 
and $q^{\Le(\R)}$
are quark fields of left (right) chirality.
The crossing terms of $\Phi^\Le$ and $\Phi^\R$ are allowed by symmetries, 
but are suppressed at high densities \cite{Son:1999cm}.
Here we take $\Phi^\Le = - \Phi^\R \equiv \Phi$ so that the ground state is positive parity state.
In the following we neglect the effect of $\U(1)_{\rm EM}$ electromagnetism since the mixing
between broken $\SU(3)_\C$ color and $\U(1)_{\rm EM}$ is sufficiently small at
high densities.
We also consider a sufficiently high density region where the masses of
three light quarks can be neglected. (We refer the reader to \cite{Eto:2009tr} 
for a recent discussion on the effect of a strange quark mass.)

In this case the GL Lagrangian for the CFL phase is given
by \cite{Iida:2000ha,Giannakis:2001wz}
\begin{eqnarray}
  \Lcal(x) 
  &=&
  \frac{\epsilon}{2} \left(\textbf{E}^a \right)^2 
  - \frac{1}{2\lambda}\left(\textbf{B}^a\right)^2
  + K_0 \Tr \left[\left(D_0 \Phi\right)^\+ D^0 \Phi \right]
  + K_1 \Tr \left[\left(D_i \Phi\right)^\+ D^i \Phi \right]
  + iK^\prime_0 \Tr \left[ \Phi^\+ D_0 \Phi\right]
  \nonumber
  \\ 
  && - V(\Phi), 
  \nonumber
  \\
  V(\Phi) &=& \Tr \left[
  \lambda_1 (\Phi^\+ \Phi)^2 - \lambda_2 \Phi^\+ \Phi 
  \right]
  + \lambda_3\left(\Tr[\Phi^\+ \Phi]\right)^2,
  \label{eq:cfl-lagrangian-0} 
\end{eqnarray}
where
$E^a_i = F^a_{0i}$, 
$B^a_i = \frac{1}{2}\epsilon_{ijk}F^a_{jk}$, 
$
F^a_{\mu\nu} = \p_\mu A^a_\nu - \p_\nu A^a_ \mu + g
  f^{abc}A^b_\mu A^c_\nu
$
,
$
 D_\mu \Phi = \left( \p_\mu - igA_\mu^aT^a \right)\Phi
$ ($i=1, \cdots ,3$),  
and $T^a$ are generators of $\SU(N)_\C$ normalized as $\Tr [T^aT^b]
=\frac{1}{2}\delta^{ab}$ with color indices $a=1,2,\cdots,N^2-1$.
Here the Lagrangian describes the low-energy effective theory in 
the CFL phase only for $N=3$, but we have extended the order parameter field 
$\Phi$ to an $N$ by $N$ matrix.
Coefficients $K_0$, $K^\prime_0$, $K_1$, $ \lambda_1$, $ \lambda_2$, and $\lambda_3$ 
are GL parameters dependent on the temperature and 
the chemical potential of the system but are dealt with as constant parameters 
in this paper. 
$\epsilon$ and $\lambda$ are the dielectric constant and the magnetic
permeability.
The form of the kinetic term of gauge fields in the effective Lagrangian
is deduced by requiring the gauge invariance, rotational invariance and
parity conservation. 
The Lorentz symmetry does not have to be maintained in general since 
superconducting matter exists.
However, there exists a modified Lorentz symmetry in which the speed of
light is replaced by $1/\sqrt{\epsilon \lambda}$.
It is always possible to restore the Lorentz invariance of the
kinetic term of gauge fields by rescaling $x^0$, $A^a_0$, $K_0$, 
$K^\prime_0$ and $K_1$.
Therefore we can start with the Lagrangian in which $\epsilon$ and $\lambda$
are taken to be unity.

For notational convenience, we introduce a vector $K_\mu \equiv
(K_0, K_1,K_1,K_1)$.
Our starting point is the following GL Lagrangian
\begin{eqnarray}
  \Lcal(x) &=& -\frac{1}{4} \left(F^a_{\mu\nu} \right)^2 
  + K_\mu \Tr \left[\left(D_\mu \Phi\right)^\+ D^\mu \Phi \right]
  + iK^\prime_0 \Tr \left[ \Phi^\+ D_0 \Phi\right]
  - V(\Phi)
  \label{eq:cfl-lagrangian}.
\end{eqnarray}
We consider the parameter region
 $\lambda_1 > 0$, 
 $\lambda_2 > 0$, 
 $\lambda_1 + N \lambda_3 > 0$.
This Lagrangian includes the most general terms which 
are consistent with the symmetry group $G$ of generalized QCD with $N$
colors and $N$ flavors,
\begin{equation}
  G = \frac{\SU(N)_\C \times \SU(N)_\F \times \U(1)_\B } 
  { ({\mathbb Z}_N)_{\C+\B} \times ({\mathbb Z}_N)_{\F+\B} },
\end{equation}
where $\SU(N)_\C$ is the local color symmetry, 
$\SU(N)_\F$ is the global flavor symmetry, 
and $\U(1)_\B$ is the global symmetry associated with the baryon number
conservation. 
Under the action of the element 
$
(V_{\C}, V_{\F}, \e^{i\theta}) \in G
$, $\Phi$ transforms as
\begin{equation}
  \Phi \rightarrow \Phi^\prime = {\rm{e}}^{i\theta}  V_{\C} \Phi V^T_{\F},
\end{equation}
where $V_\C\in \SU(N)_\C$, $V_\F \in \SU(N)_\F, \e^{i\theta} \in \U(1)_\B$.
The elements corresponding to the discrete groups in the denominator of $G$
can be written as 
$
(z_1, z_2, z_1^{-1} z_2^{-1}) \in G
$
with $z_1, z_2 \in {\mathbb Z}_N$.
These transformations are removed since they do not change $\Phi$ for any value of $\Phi$.

In the CFL phase the free energy is minimized when
$\Phi$ is proportional to a constant unitary matrix.
By using the symmetry $G$, one can take the value of $\Phi$ 
without loss of generality as
\begin{equation}
  \Phi = |\Delta| \mathbf{1}_N,
\end{equation}
where $|\Delta|$ is a real number. 
By this expectation value, 
$\Phi$ is invariant under the restricted transformations 
\begin{equation}
  \{h | h = (Uz^{-1}, U^{\ast}, z), U \in \SU(N), z \in {\mathbb Z}_N \} \subset G.
\end{equation}
Therefore the symmetry $G$ is spontaneously broken 
in the ground state to
\begin{equation}
  H = \frac{\SU(N)_{\C+\F}
  }
  {({\mathbb Z}_N)_{\C+\F}
  }.
\end{equation}
The order parameter space which characterizes the 
degenerate ground states is given by
\begin{equation}
  G / H \simeq \frac{ \U(1) \times \SU(N)}{{\mathbb Z}_N} \simeq \U(N).
\end{equation} 
The CFL phase admits stable vortices 
since the first homotopy group of the order parameter space is nontrivial: 
\begin{equation}
  \pi_1 \left( G/H \right) \simeq {\mathbb  Z}.
\end{equation}

\section{The dual transformation}\label{sec:dual}

In this section, we perform a dual transformation within path integral 
to derive a dual Lagrangian for the CFL phase.
After the transformation, massive gluons are described by massive
non-Abelian antisymmetric tensor fields \cite{Seo:1979id} and $\U(1)_\B$ phonons 
are described by massless antisymmetric tensor fields.
We show that in the dual description vortices appear as sources which
can absorb or emit these particles.
This is consistent with the empirical rule that a dual
transformation interchanges the role of particles and solitons.

\subsection{The dual transformation of massive gluons}

The partition function of the CFL phase can be written as
\begin{equation}
  Z = \int \Dcal A^a_\mu(x) \Dcal \Phi(x) \exp \left\{ i\int
  d^4x \Lcal(x)  \right\},
  \label{partition-function}
\end{equation}
with the Lagrangian defined in Eq.~(\ref{eq:cfl-lagrangian}).
We shall impose the gauge fixing condition on the field $\Phi$ rather than
on the gauge fields since they are integrated out in the end.
The gauge fixing condition is taken care of 
when we consider a concrete vortex solution.

Let us introduce non-Abelian antisymmetric tensor fields $B^a_{\mu\nu}$ by 
a Hubbard-Stratonovich transformation
\begin{equation}
  \exp \left[ i\int \diff^4 x \left\{ -\frac{1}{4}(F^a_{\mu\nu})^2
				  \right\} \right]
  \propto \int \Dcal B^a_{\mu\nu}\exp \left[ i\int \diff^4x
  \left\{ -\frac{1}{4} \left[ m^2 (B^a_{\mu\nu})^2 - 2m \tilde{B}^a_{\mu\nu}
		      F^{a,\mu\nu}) \right]  \right\}
  \right],
  \label{dual-transformation}
\end{equation}
where 
$\tilde{B}^a_{\mu\nu} \equiv \frac{1}{2} \epsilon_{\mu\nu\rho\sigma} B^{a, \rho \sigma}$. 
The parameter $m$ introduced above is a free parameter at this stage.
We will choose $m$ later so that the kinetic term of $B^a_{\mu\nu}$ is
canonically normalized. 

Substituting (\ref{dual-transformation}) into (\ref{partition-function}), 
we can now perform the integration over the gauge fields $A^a_\mu$.
The degrees of freedom of gluons are expressed by $B^a_{\mu\nu}$
after this transformation.
Each term in the Lagrangian is transformed as follows:
\begin{eqnarray}
  && K_\mu \Tr\{ (D_\mu \Phi)^\+(D^\mu \Phi)\} 
  + iK^\prime_0 \Tr \left[\Phi^\+ D_0 \Phi\right] 
  \nonumber\\
  &&= K_\mu \Tr \left\{ \Phi^\+ (\overleftarrow{\p}_\mu + igA^a_\mu T^a )
  (\overrightarrow{\p}^\mu - igA^{b,\mu} T^b) \Phi \right\} 
  + iK^\prime_0 \Tr \left[\Phi^\+ (\p_0 -ig A^a_\mu T^a) \Phi\right] 
  \nonumber\\
  &&=K_\mu \Tr \{(\p_\mu\Phi)^\+(\p^\mu\Phi)\}  
  + iK^\prime_0 \Tr \left[\Phi^\+ \p_0 \Phi\right]  
  + gA^a_\mu J^{a,\mu} \nonumber\\
  && \quad + g^2 g_{\mu\nu} \sqrt{K_\mu K_\nu} A^{a,\mu} A^{b,\nu}  
  \Tr \left[ \Phi^ \+ T^aT^b \Phi \right],
\end{eqnarray}
with $ 
J^a_\mu \equiv -iK_\mu \Tr \left[ \Phi^\+ (\overleftarrow{\p}_\mu -
		   \overrightarrow{\p}_\mu)T^a \Phi  \right]
- K^\prime_0 \Tr \left[\Phi^\+  T^a \Phi\right]
$, and
\begin{equation}
 \begin{split}
   -\frac{1}{2}m \tilde{B}^a_{\mu\nu} F^{a,\mu\nu} &= -\frac{1}{2}m
   \tilde{B}^a_{\mu\nu} ( 2\p_\nu A^a_\mu+g f^{abc}A^b_\mu A^c_\nu )\\
   &= m A^a_\mu \p_\nu \tilde{B}^a_{\mu\nu} + \frac{1}{2}mgf^{abc}
   A^a_\mu A^b_\nu \tilde{B}^c_{\mu\nu} .
 \end{split}
\end{equation}
Performing the integration over $A^a_\mu$, the following part
of the partition function is rewritten as
\begin{equation}
 \begin{split}
  & \int \Dcal A^a_\mu \exp \left\{ i \int \diff^4 x \left[ \frac{1}{2}g^2 A^{a,\mu}
  K^{ab}_{\mu\nu} A^{b,\nu} - m\left( \p^\nu \tilde{B}^a_{\mu\nu}
  - \frac{g}{m}J^a_\mu \right) A^{a,\mu} \right] \right\} 
  \\
  & \propto (\det K^{ab}_{\mu\nu})^{-1/2} \exp \left\{
  i\int \diff^4 x \left[ -\frac{1}{2} \left(\frac{m}{g} \right)^2
  \left(\p_\rho \tilde{B}^{a,\mu\rho} - \frac{g}{m}J^{a,\mu} \right)
  \left(K^{-1} \right)^{ab}_{\mu\nu}
  \left(\p_\sigma \tilde{B}^{b,\nu\sigma} - \frac{g}{m}J^{b,\nu} \right)
  \right] \right\},
 \end{split}
\end{equation}
where $K^{ab}_{\mu\nu}$ is defined by
\begin{equation}
  \begin{split}
    K^{ab}_{\mu\nu} &= \frac{1}{2} g_{\mu\nu} \sqrt{K_\mu K_\nu} \Tr \left[\Phi^\+ T^aT^b \Phi \right] -
    \frac{m}{g}f^{abc} \tilde{B}^c_{\mu\nu} 
    \\
    & \equiv \bm{\Phi}^{ab}_{\mu\nu} - \frac{m}{g} \hat{B}^{ab}_{\mu\nu},
  \end{split}
\end{equation}
with
$
\bm{\Phi}^{ab}_{\mu\nu} \equiv \frac{1}{2} g_{\mu\nu} \sqrt{K_\mu K_\nu} \Tr
\left[\Phi^\+ T^aT^b \Phi \right]
$ and 
$
\hat{B}^{ab}_{\mu\nu} \equiv f^{abc} \tilde{B}^c_{\mu\nu}
$. We define the inverse of $K^{ab}_{\mu\nu}$ by the power-series expansion in $1/g$
\begin{equation}
 \begin{split}
  K^{-1} 
  = \left(\bm{\Phi} - \frac{m}{g}\hat{B} \right)^{-1} 
  = \bm{\Phi}^{-1} \sum_{n=0}^\infty \left( \frac{m}{g} \hat{B}
  \bm{\Phi}^{-1}\right)^n.
 \end{split}
\end{equation}
As a result, we obtain the following partition function
\begin{equation}
  Z \propto
  \int \Dcal B^a_{\mu\nu}(\det K^{ab}_{\mu\nu})^{-1/2} \exp \left\{
  i\int \diff^4 x \Lcal^\ast_{\rm G}(x) \right\}
  ,
\end{equation}
where $\Lcal^\ast_{\rm G}$ denotes the gluonic part of the dual Lagrangian 
\begin{equation}
  \Lcal^\ast_{\rm G} =  -\frac{1}{2} \left(\frac{m}{g} \right)^2
  \left(\p_\rho \tilde{B}^{a,\mu\rho} - \frac{g}{m}J^{a,\mu} \right)
  \left(K^{-1} \right)^{ab}_{\mu\nu}
  \left(\p_\sigma \tilde{B}^{b,\nu\sigma} - \frac{g}{m}J^{b,\nu} \right)
  -\frac{1}{4}m^2(B^a_{\mu\nu})^2
  .
  \label{eq:dual-lagrangian}
\end{equation}

Now we define the non-Abelian vorticity tensor $\omega^a_{\mu\nu}$ 
as the coefficient of the term linearly
proportional to $B^a_{\mu\nu}$. Collecting relevant terms in the above
Lagrangian, the coupling between massive gluons and 
the vorticity is given by
\begin{eqnarray}
  \Lcal^{\ast}_{\rm G} 
  &\supset&
  \frac{1}{2} \frac{m}{g} \left[ 
  \p_\rho \tilde{B}^{a,\mu\rho} (\bm{\Phi}^{-1})^{ab}_{\mu\nu}J^{b,\nu} +
  J^{a,\mu} (\bm{\Phi}^{-1})^{ab}_{\mu \nu} \p_\rho \tilde{B}^{b,\nu\rho}
  \right]
  -\frac{1}{2}\left(\frac{m}{g}\right) J^{a,\mu}[\bm{\Phi}^{-1}\hat{B}\bm{\Phi}^{-1}]^{ab}_{\mu\nu}J^{b,\nu} 
  \nonumber \\
  & \equiv& -\frac{1}{2} \left( \frac{m}{g} \right)B^a_{\lambda \sigma}
  \omega^{a,\lambda \sigma},
\end{eqnarray}
where we have defined the vorticity tensor $\omega^a_{\mu\nu}$ as 
\begin{equation}
  \omega^{a,\lambda \sigma}
  \equiv \epsilon^{\lambda\sigma\mu\nu}\left[ \p_\nu \left\{
  (\bm{\Phi}^{-1})(^{ab}_{\mu\rho})
  J^{b,\rho} \right\} +
  J^{e,\alpha}(\bm{\Phi}^{-1})^{ec}_{\alpha\mu}f^{cda}(\bm{\Phi}^{-1})^{db}_{\nu\beta}J^{b,\beta}
				     \right].
  \label{eq:vorticity}
\end{equation}
Here $A{(^{ab}_{\mu\nu})}$ is a symmetrized summation defined by
$A{(^{ab}_{\mu\nu})}\equiv A^{ab}_{\mu\nu}+A^{ba}_{\nu\mu}$.
This expression for the non-Abelian vorticity is valid for general
vortex configurations.
The information of vortex configuration is included in $\bm{\Phi}$ and $J^a_\mu$.

\subsection{The dual transformation of $\U(1)_\B$ phonons}

In the following, we perform a dual transformation of the NG boson 
associated with the breaking of $\U(1)_\B$ symmetry.
This mode corresponds to the fluctuation of the overall phase of $\Phi$ 
which can be parametrized as $\Phi(x) = \e^{i\pi(x)}\psi(x)$, 
where $\pi(x)$ is a real scalar field.
Substituting this into the following part in the Lagrangian
(\ref{eq:cfl-lagrangian}) leads to \footnote{
The term $\Tr \left[\p_\mu \psi^\+ \psi -  \psi^\+ \p_\mu \psi \right]$
automatically vanishes since $\psi$ can be decomposed as $\psi = (\Delta+\rho)
\bm{1}_N + (\chi^a + i \zeta^a  ) T^a
$
and the modes $\zeta^a$ are absorbed by gluons.
}
\begin{equation}
 \begin{split}
  K_\mu  \Tr\{ (\p_\mu \Phi)^\+ (\p^\mu \Phi) \} 
  + iK^\prime_0 \Tr \{ \Phi^\+ \p_0 \Phi \} 
  & =
  K_\mu \left(\p_\mu \pi \right)^2 M^2
  - \p^\mu \pi J^0_\mu + K_\mu \Tr (\p_\mu \psi)^2
  +i K^\prime_0 \Tr \{ \psi^\+ \p_0 \psi \}
  ,
 \end{split}
\end{equation}
with 
$
J^0_\mu \equiv
\delta_{\mu 0} K^\prime_0M^2
$
and $M^2 \equiv \Tr\left[\psi^\+ \psi\right]$.
We will transform the $\U(1)_\B$ phonon field $\pi(x)$ into a massless two-form
field $B^0_{\mu\nu}$. 
Note that the field $\pi(x)$ has a multivalued part in general; 
since $\pi(x)$ is the phase degree of freedom, 
$\pi(x)$ can be multivalued without violating the
single-valuedness of $\Phi(x)$. 
In fact the multivalued part of $\pi(x)$
corresponds to a vortex. 
Let us denote the multivalued part of $\pi(x)$ as $\pi_{MV}(x)$. 

The dual transformation of this $\U(1)_{\B}$ phonon field is essentially 
the same as the case of a superfluid.
We basically follow the argument of \cite{Lee:1993ty}.
Let us introduce an auxiliary field $C_\mu$ 
by linearizing the kinetic term of $\pi(x)$ 
in the partition function as follows
\begin{equation}
 \begin{split}
  Z & \propto 
  \int \Dcal \pi \Dcal \pi_{MV} \exp i
  \left[
  \int \diff^4x \left(
   M^2 K_\mu \left\{\p_\mu (\pi + \pi_{MV}) \right\}^2
  - \p^\mu (\pi + \pi_{MV}) J^0_\mu
  \right)
  \right]  
  \\
  & \propto 
  \int \Dcal \pi \Dcal \pi_{MV} \Dcal C_\mu \exp i
  \left[
  \int \diff^4x \left(
    -\frac{C^2_\mu}{M^2} - 2 C_\mu \sqrt{K_\mu} \p^\mu (\pi+\pi_{MV})
    - \p^\mu(\pi + \pi_{MV}) J^0_\mu
  \right)
  \right] .
 \end{split}
\end{equation}
Integration over $\pi(x)$ gives a delta function
\begin{equation}
  \int \Dcal \pi 
  \exp i
  \left[
  \int \diff^4x \left(
   -2  C_\mu \sqrt{K_\mu} \p^\mu \pi
   +\pi \p^\mu J^0_\mu
  \right)
  \right] 
  =  
  \delta
  \left\{
  \p^\mu \left(
   2  C_\mu \sqrt{K_\mu}
   +  J^0_\mu
  \right)
  \right\}.
\end{equation}
Then let us introduce the dual antisymmetric tensor field $B_{\mu\nu}^0$ by
\begin{equation}
  \int \Dcal C_\mu   
  \delta
    \left\{
     \p^\mu \left(
      2  C_\mu\sqrt{K_\mu} +  J^0_\mu
    \right)
  \right\} \cdots 
  = 
  \int \Dcal C_\mu \Dcal B_{\mu\nu}^0
  \delta
    \left(
      2  C_\mu \sqrt{K_\mu} +  J^0_\mu - m^0 \p^\nu \tilde{B}^0_{\mu\nu}
  \right) \cdots
\end{equation}
where the dots denote the rest of the integrand and 
$m^0$ is a parameter.
By this change of variables we have introduced an infinite gauge volume, 
corresponding to the transformation $\delta B^0_{\mu\nu}  = \p_\mu
\Lambda_\nu - \p_\nu \Lambda_\mu$ with a massless vector field $\Lambda_\mu$.
This can be taken care of by fixing the gauge later.
There is no nontrivial Jacobian factor as the change of
variables is linear.
Integrating over $C_\mu$, and transforming a resultant term in the Lagrangian as
\begin{equation}
  \begin{split} 
   m^0 \p^\nu \tilde{B}^0_{\mu\nu} \p^\mu \pi_{MV} &= 
   -m^0 {B}^{0,\rho\sigma} \epsilon_{\mu\nu\rho\sigma} \p^\nu \p^\mu \pi_{MV} 
   \\
   & \equiv 
   - 2\pi m^0 B^{0,\rho\sigma}\omega^0_{\rho\sigma} ,
  \end{split}
\end{equation}
where the first equality holds up to a total derivative and 
we have defined
\beq
  \omega^0_{\rho\sigma} \equiv \frac{1}{2\pi}\epsilon_{\mu\nu\rho\sigma} \p^\nu \p^\mu
  \pi_{MV}\,.
\eeq
We thus obtain the dual Lagrangian for the $\U(1)_\B$ phonon part
\begin{equation}
  \begin{split}
   \Lcal^\ast_{\rm Ph}&= 
    -\left(\frac{1}{2M}\right)^2 K_\mu
     ( m^0 \p_\nu \tilde{B}^0_{\mu\nu} - J^0_\mu )^2
   -2\pi m^0 B^{0,\mu\nu} \omega^0_{\mu\nu} 
   \label{eq:Ph}.
  \end{split}
\end{equation}
Note that the term linear in $B^0_{\mu\nu}$ coming from the first term of
(\ref{eq:Ph}) is a total derivative and does not contribute to the
equation of motion. 
The partition function is proportional to 
\begin{equation}
  Z \propto \int \Dcal \pi_{MV} \Dcal B^0_{\mu\nu} \exp i\left[\int \diff^4
  x \Lcal^\ast_{\rm Ph}\right].
\end{equation}
The $\U(1)_\B$ phonons are now described by a massless two-form field 
$B^0_{\mu\nu}$ and vortices appear as sources for $B^0_{\mu\nu}$.

\subsection{The dual Lagrangian}

We now summarize the results obtained so far.
We have shown that the partition function $Z$ of the CFL phase 
is proportional to $Z^\ast$ with the dual Lagrangian $\Lcal^\ast$:
\begin{equation}
  Z \propto Z^\ast = 
  \int \Dcal B^a_{\mu\nu} \Dcal \pi_{MV} \Dcal B^0_{\mu\nu} \Dcal \psi~
  (\det K^{ab}_{\mu\nu})^{-1/2} \exp \left\{
  i\int \diff^4 x \Lcal^\ast (x) \right\},
\end{equation}
where
\begin{equation}
  \Lcal^\ast = \Lcal^\ast_{\rm G} + \Lcal^\ast_{\rm Ph} + K_\mu \Tr
  (\p_\mu \psi)^2 
  + iK^\prime_0 \Tr \{\psi^\+ \p_0 \psi\} 
  - V(\psi) .
\end{equation}
Here $\Lcal^\ast_{\rm G}$ and $\Lcal^\ast_{\rm Ph}$ are given 
in (\ref{eq:dual-lagrangian}) and (\ref{eq:Ph}), respectively.
We can discuss the interaction between vortices and quasiparticles
in terms of the dual Lagrangian.
Vortices are expected to appear as a source term for gluons and $\U(1)_\B$ phonons.
The result above is valid for general vortex configurations.

\section{Coupling of non-Abelian vortex with 
dual fields}\label{sec:vor-tensor}

In the last section we have obtained the dual Lagrangian for general
vortex configurations.
In order to see the coupling of quasiparticles with internal
orientational degrees of freedom explicitly 
and to discuss physical consequences, let us consider a single-vortex solution 
and find an expression of the vorticity tensor for this case.

\subsection{Vortex solution}

We consider a cylindrically symmetric vortex configuration along the $z$-axis.
Then, the profile of a vortex solution with the lowest energy takes the form
\cite{Balachandran:2005ev,Eto:2009kg}
\begin{equation}
 \Phi_0(x)=
  \begin{pmatrix}
   f(r)\e^{i\theta}  &0 \\
   0 & g(r)\bm{1}_{N-1} \\
  \end{pmatrix},
\end{equation}
where $f(r)$ and $g(r)$ are functions of the radial coordinate which can be
determined by solving equations of motions 
under their asymptotic behaviors
$
(f,g) = (|\Delta|, |\Delta|) 
$ 
as 
$
r \rightarrow \infty
$
and
$
(f,g^\prime) = (0, 0) 
$ 
as 
$
r \rightarrow 0.
$

Let us identify the $\mathbb{C}P^{N-1}$ zero mode in the background
solution.
To this end, it is convenient to take the singular gauge.
We perform a gauge transformation on this solution
by $V \in \SU(N)_\C$, 
\begin{equation}
  V=
  \begin{pmatrix}
   \e^{-i\theta(N-1)/N}  &0 \\
   0 & \e^{i\theta/N} \bm{1}_{N-1} \\
  \end{pmatrix}\,,
\end{equation}
which transforms $\Phi$ as 
\begin{equation}
 \begin{split}
  \Phi^\ast&=V\Phi_0
  \\
  &=
  \e^{i\theta/N}
  \begin{pmatrix}
   f(r)  &0 \\
   0 & g(r)\bm{1}_{N-1} \\
  \end{pmatrix}
  \\
  & \equiv \e^{i\theta/N}
  \left\{F(r)\bm{1}_{N}+G(r)
  \begin{pmatrix}
   -\frac{N-1}{N} & 0 \\
   0 & \frac{1}{N} \bm{1}_{N-1}
  \end{pmatrix}
  \right\},
 \end{split}
\end{equation}
where $F(r)$ and $G(r)$ are functions of the radial coordinate which are
related to $f(r)$ and $g(r)$ as 
\begin{eqnarray}
   f =  F - \frac{N-1}{N}G, \quad
   g =  F + \frac{1}{N}G.
\end{eqnarray}
Under this gauge transformation, 
the vortex solution is physically unchanged. 
At this stage we fix the gauge by fixing all the local color
transformation of $\Phi$ \cite{Eto:2009bh}.

General vortex solutions are obtained by acting color-flavor locked
 $\SU(N)_{\C+\F}$ transformations on $\Phi^\ast$.
Because of the existence of the vortex, the symmetry further breaks down to
$\SU(N-1) \times \U(1)$.
We parametrize the associated $\mathbb{C}P^{N-1}$ orientational moduli space,
which is obtained by performing the $\SU(N)_{\C+\F}$ transformation on
 $\Phi^\ast$ as 
\begin{equation}
  \Phi = U \Phi^\ast U^{-1}
  = \e^{i\theta/N}\left\{F(r)\bm{1}_{N}+G(r)
  \left( \phi \phi^\+ - \frac{1}{N}\bm{1}_{N} \right) \right\}
  ,
\end{equation}
where $U \in \SU(N)$, $\phi$ is a complex $N$-component vector
which transforms as the fundamental representation of
$\SU(N)_{\C+\F}$ and satisfies the relation $\phi^\+\phi=1$.

The definition of $\phi$ has a redundancy in the overall phase.
We cannot distinguish $\phi$ and $\e^{i\alpha}\phi$ 
because both express the same solution, 
so they should identified: $\phi \simeq \e^{i\alpha}\phi$.
Therefore $\phi$ are indeed the homogeneous coordinates on $\mathbb{C}P^{N-1}$.
The low-energy excitation on the non-Abelian vortex can be described by
the $\mathbb{C}P^{N-1}$ model.
It has been shown that the $\mathbb{C}P^{N-1}$ modes are in fact normalizable
and localized around the vortex \cite{Eto:2009bh,Eto:2009tr}.

We shall express the vorticity tensor by orientational zero  modes $\phi$
and the profile functions.
$\bm{\Phi}^{ab}_{\mu\nu}$ can be written as
\begin{equation}
 \begin{split}
  \bm{\Phi}^{ab}_{\mu\nu} &=
  \frac{1}{2}g_{\mu\nu} \sqrt{K_\mu K_\nu} \Tr \left[ \Phi^\+ T^a T^b
  \Phi\right] 
  \\
  &=
  \frac{1}{4}C g_{\mu\nu} \sqrt{K_\mu K_\nu} \left( \delta^{ab} +
  \frac{2D}{C}\phi^\+ T^aT^b \phi \right),
 \end{split}
\end{equation}
where we have defined $C(r)$ and $D(r)$ by
\beq
  C(r) \equiv (F-G/N)^2, \quad 
  D(r) \equiv G^2 + 2G(F-G/N),
\eeq 
whose asymptotic behaviors are 
$(C, D) \simeq (|\Delta|^2, 0) $ as $r \rightarrow \infty$ and \\
$(C, D) \simeq
\left\{ (1-2/N)^2 |\Delta|^2, (1-2/N)^2 |\Delta|^2 \right\} $
as $r \rightarrow 0$. 
We define the inverse of $\bm{\Phi}^{ab}_{\mu\nu}$ 
by an expansion in the dimensionless quantity $2D/C$.
The latter is a deviation of $\Phi$ from the ground state
value and small except for the center of the vortex. 
Although we expand $(\bm{\Phi}^{-1})^{ab}_{\mu\nu}$ in power-series, 
we will sum up all the terms later.
Therefore $(\bm{\Phi}^{-1})^{ab}_{\mu\nu}$ is calculated exactly 
whenever the condition $2D/C < 1$ is satisfied.
This condition holds except for the vicinity of the vortex core. 
In matrix notation, $\Phi$ reads
\begin{equation}
  \bm{\Phi} = \frac{C}{4} \hat{K} \left( \bm{1} + \frac{2D}{C}\hat{\phi} \right),
\end{equation}
where the components of the matrix $\hat{\phi}$ and $\hat{K}$ are given by $\hat{\phi}^{ab} =
\phi^\+ T^a T^b \phi$ and $K_{\mu\nu} = g_{\mu\nu}\sqrt{K_\mu K_\nu}$\,. 
Then $(\bm{\Phi}^{-1})^{ab}_{\mu\nu}$ can be written as 
\begin{equation}
  \bm{\Phi}^{-1} = \frac{4}{C}\hat{K}^{-1} \left( \bm{1} + \frac{2D}{C}\hat{\phi}
  \right)^{-1} 
  = \frac{4}{C}\hat{K}^{-1}  \sum_{n=0}^{\infty} \left( -\frac{2D}{C}\hat{\phi}  \right)^n .
\end{equation}
More explicitly, 
\begin{equation}
  (\bm{\Phi}^{-1})^{ab}_{\mu\nu} 
  =
  g_{\mu\nu}\frac{1}{\sqrt{K_\mu K_\nu}} \frac{4}{C}
  \left(\delta^{ab} - \frac{2D}{C}\phi^\+ T^aT^b \phi 
  + \left(-\frac{2D}{C}\right)^2 \phi^\+ T^a T^c \phi
  \phi^\+ T^c T^b \phi + \cdots
  \right).
\end{equation}

On the other hand, the current $J^a_\mu$ is written as
\begin{eqnarray}
  J^a_\mu 
  &=& - 2K_\mu\Im \Tr \left[ \Phi^\+ \p_\mu T^a \Phi \right] 
  - \delta_{\mu 0} K^\prime_0 \Tr \{ \Phi^\+ T^a \Phi\}  
  \nonumber\\
  &=& -\frac{2}{N}K_\mu 
  \left\{ 
  \p_\mu \theta + \frac{NK^\prime_0}{2K_0} \delta_{\mu 0}
  \right\} 
  D(r)  \phi^\+ T^a \phi - 2G^2K_\mu \Im \Tr \left[
  \phi\phi^\+  T^a 
  \left\{ \p_\mu ( \phi\phi^\+ ) \right\}
  \right].
\end{eqnarray}

\subsection{The vorticity tensor}

Let us see how the vorticity tensor can be expressed in terms of profile 
functions and orientational zero modes in the case of a single-vortex configuration discussed in the last section. 
The Abelian component $\omega^0_{\mu\nu}$ of the vorticity tensor is
readily identified as
\begin{equation}
  \omega^0_{\mu\nu} =  \frac{1}{2\pi N}\epsilon_{\mu\nu\rho\sigma}
  \p^\sigma \p^\rho \theta.
  \label{eq:abelian-vorticity}
\end{equation}

The non-Abelian component $\omega^a_{\mu\nu}$ of the vorticity tensor is
obtained by substituting $J^a_\mu$ and $\bm{\Phi}^{-1}$ into (\ref{eq:vorticity}).
In order to simplify the following calculations, 
let us define \cite{Eto:2009bh}
\beq
  {\cal F}_\mu(a,b) \equiv 
  a \phi \p_\mu{\phi}^\+ + b \p_\mu\phi\phi^\+ 
  + (a-b) \phi\phi^\+ \p_\mu \phi\phi^\+,\quad  a,b \in \mathbb{C}.
\eeq
This quantity ${\cal F}$ satisfies the following relations
\beq
 &&
 {\cal F}_\mu(a,b)^\+ = {\cal F}_\mu(b^*,a^*),\quad  \Tr [{\cal F}_\mu(a,b)]=0,
 \\
 &&
 \alpha {\cal F}_\mu(a,b) = {\cal F}_\mu(\alpha a,\alpha b), \quad
    {\cal F}_\mu(a,b) + {\cal F}_\mu(a',b') = {\cal F}_\mu(a+a',b+b') ,
 \\ 
 &&
 \phi^\+ {\cal F}_\mu(a,b) \phi = 0 , \quad
 \phi^\+ \left[ {\cal F}_\mu(a,b), T^a  \right] \phi = \Tr \left( {\cal
 F}_\mu(a,-b)T^a \right) ,\\
 &&
 \Tr \left\{T^a[{\cal F}_{[\mu}(a,b)^\+,{\cal F}_{\nu]}(a,b) ]
 \right\} 
 \\
 &&
 \quad= 
 (|a|^2 + |b|^2) \left\{
 \phi^\+ T^a \phi \p_{[\nu} \phi^\+ \p_{\mu]}\phi + \p_{[\nu}\phi^\+ T^a
 \phi_{\mu]}\phi
 + 
 \phi^\+ T^a \p_{[\nu}\phi \phi^\+ \p_{\mu]}\phi 
 + \phi^\+ \p_{[\nu}\phi \phi^\+ T^a \p_{\mu]} \phi
 \right\} ,
 \nonumber\\
\eeq
where $[...]$ denotes antisymmetrization of indices.
The current $J^a_\mu$ for $a = 1, \dots, N^2-1$ can be written in terms
of ${\cal F}$ as
\begin{equation}
 \begin{split}
  J^a_\mu &= 
  - \frac{2}{N} D K_\mu 
  \left\{ 
  \p_\mu \theta + \frac{NK^\prime_0}{2K_0} \delta_{\mu 0}
  \right\}
  \phi^\+ T^a \phi 
  -2G^2K_\mu \Im \Tr \left\{
		 \phi\phi^\+ T^a \p_\mu (\phi\phi^\+)
		\right\} \\
  &=
  - \frac{2}{N} D K_\mu 
  \left\{ 
  \p_\mu \theta + \frac{NK^\prime_0}{2K_0} \delta_{\mu 0}
  \right\}
  \phi^\+ T^a \phi 
  + 
  iG^2K_\mu \Tr \left\{ T^a {\cal F}_\mu(-1,1)\right\} 
  \\
  & \equiv (J_1)^a_\mu + (J_2)^a_\mu .
  \label{eq:current-with-f}
 \end{split}
\end{equation}

As is shown in the appendix, 
the following quantities can be written in terms of ${\cal F}$ as
\beq
  (\bm{\Phi}^{-1}J)^a_\mu & = -\frac{8}{N} \gamma \phi^\+T^a\phi  
  \left\{ 
  \p_\mu \theta + \frac{NK^\prime_0}{2K_0} \delta_{\mu 0}
  \right\} 
  + i \alpha \Tr \left[ T^a {\cal F}_\mu( -1, \beta ) \right]
  \label{eq:phi-j} 
  \nonumber\\
  & \equiv (\bm{\Phi}^{-1}J_1)^a_\mu + (\bm{\Phi}^{-1}J_2)^a_\mu,
\eeq
\beq
  (J\bm{\Phi}^{-1})^a_\mu = -\frac{8}{N} \gamma \phi^\+ T^a \phi 
  \left\{ 
  \p_\mu \theta + \frac{NK^\prime_0}{2K_0} \delta_{\mu 0}
  \right\}
  -i \alpha \Tr \left[ T^a {\cal F}_\mu( \beta,- 1 ) \right],
  \label{eq:j-phi}
\eeq
where the functions $\alpha(r)$, $\beta(r)$ and $\gamma(r)$ are defined by
\beq
  \alpha(r) \equiv \frac{4G^2}{C}\,, \quad
  \beta(r) \equiv \frac{C}{C+D}\,,  \quad
  \gamma(r) \equiv \frac{D}{C+D(1-1/N)}\,.
\eeq
Hence the following part in the first term of the vorticity tensor reads
\beq
  (J\bm{\Phi}^{-1}+\bm{\Phi}^{-1}J)^a_\mu 
  = 
  -\frac{16}{N} \gamma \phi^\+ T^a \phi 
  \left\{ 
  \p_\mu \theta + \frac{NK^\prime_0}{2K_0} \delta_{\mu 0}
  \right\} 
  + i \alpha (1+\beta) \Tr \left[ T^a {\cal F}_\mu(-1, 1) \right],
\eeq
where we have used the linearity of ${\cal F}$. 
The second term in (\ref{eq:vorticity}) is also expressed by ${\cal F}$, 
which is proportional to
\begin{equation}
 \begin{split}
  (J\bm{\Phi}^{-1})^b_\mu f^{abc} (\bm{\Phi}^{-1}J)^c_{\nu} &=  
  ((J_1+J_2)\bm{\Phi}^{-1})^b_\mu f^{abc}
  (\bm{\Phi}^{-1}(J_1+J_2))^c_{\nu} 
  \\
  &=(J_1\bm{\Phi}^{-1} f \bm{\Phi}^{-1}J_2)^a_{\mu\nu}
  +(J_2\bm{\Phi}^{-1} f \bm{\Phi}^{-1}J_1)^a_{\mu\nu}
  +(J_2\bm{\Phi}^{-1} f \bm{\Phi}^{-1}J_2)^a_{\mu\nu}\,.
 \end{split}
\end{equation}
Each term is calculated as follows:
\begin{eqnarray}
  && (J_1\bm{\Phi}^{-1} f \bm{\Phi}^{-1}J_2)^a_{\mu\nu}
  +(J_2\bm{\Phi}^{-1} f \bm{\Phi}^{-1}J_1)^a_{\mu\nu} 
  \nonumber\\
  &&= 
  -\frac{8i}{N} \gamma \phi^\+ T^b \phi 
  \left\{ 
  \p_\mu \theta + \frac{NK^\prime_0}{2K_0} \delta_{\mu 0}
  \right\}
  f^{abc} \Tr (T^c {\cal
  F}_\nu(-\alpha, \alpha \beta)) + \cdots
  \nonumber\\
  &&= \frac{4}{N} \alpha \gamma (1+\beta) 
  \left\{ 
  \p_\mu \theta + \frac{NK^\prime_0}{2K_0} \delta_{\mu 0}
  \right\} 
  \Tr \left\{
  {\cal F}_\nu(1, 1)T^a
  \right\},
\end{eqnarray}
\begin{eqnarray}
  &&(J_2\bm{\Phi}^{-1})^b_\mu f^{abc} (\bm{\Phi}^{-1}J_2)^c_{\nu} 
  \nonumber\\
  &&=
  - \Tr \left[ T^b {\cal F}_\mu( -\alpha \beta, \alpha) \right] f^{abc}
  \Tr \left[ T^c {\cal F}_\nu( -\alpha, \alpha \beta ) \right] 
  \nonumber\\
  &&=
  -\frac{i\alpha^2}{2}
  \Tr 
  \left\{ 
   T^a \left[ {\cal F}_\mu(-1, \beta)^\+, {\cal F}_\nu(-1, \beta) \right]
  \right\},
\end{eqnarray}
where we have used $f^{abc}T^c =-i[T^a,T^b]$ 
and traceless property of ${\cal F}$.
This quantity is explicitly rewritten in terms of $\phi$ as
\begin{align}
  & \Tr \left\{ T^a \left[ {\cal F}_{[\mu}(-1, \beta )^\+, {\cal
  F}_{\nu]}( -1, \beta)
  \right]\right\}
  \nonumber \\
  & = 
  (1+\beta^2)
  \left[
   \phi^\+ T^a \phi \p_{[ \mu } \phi^\+  \p_{ \nu] }
  \phi
   +  \p_{[ \mu } \phi^\+  T^a  \p_{ \nu] } \phi
  + \phi^\+ T^a \p_{[\mu} \phi \p_{\nu]}\phi^\+ \phi
  + \p_{[\mu} \phi^\+ T^a \phi \p_{\nu]}\phi^\+ \phi
  \right], 
\end{align} 
where we have antisymmetrized the quantity with respect to $(\mu,
\nu)$, because symmetric part vanishes when contracted with a completely
  antisymmetric tensor $\epsilon_{\lambda\sigma\mu\nu}$.
Therefore non-Abelian components of the vorticity tensor is written
in terms of ${\cal F}$ as 
\begin{equation}
 \begin{split}
  \omega^a_{\lambda \sigma} 
  =
  \epsilon_{\lambda \sigma \mu \nu} & \left[ 
  \p^\nu 
  \left\{
  -\frac{16}{N} \gamma \phi^\+ T^a \phi 
  \left(
  \p^\mu \theta + \frac{NK^\prime_0}{2K_0} \delta^{\mu 0}
  \right) 
  +i\alpha(1+\beta) \Tr \left[ T^a {\cal F}^\mu(-1, 1) \right] 
  \right\}
  \right. 
  \\
  &
  \left.
  +
  \frac{4}{N} \alpha \gamma (1+\beta) 
  \left(
  \p^{[\mu} \theta + \frac{NK^\prime_0}{2K_0} \delta^{[\mu 0}
  \right) 
  \Tr \left[ T^a {\cal F}^{\nu]}(1,1) \right]  
  \right.
  \\
  &
  \left.
  -
  \frac{i\alpha^2}{2}
  \Tr 
  \left\{ 
  T^a \left[ {\cal F}^{[\mu}( -1, \beta)^\+ , {\cal F}^{\nu]}( -1, \beta) \right]
  \right\}
  \right]. 
 \end{split} 
\end{equation}
This can be represented explicitly by orientational zero modes as 
\begin{multline}
  \omega^a_{\lambda \sigma} = \epsilon_{\lambda \sigma \mu \nu} 
  \left[
  \p^\nu 
  \Big\{
  -\frac{16}{N} \gamma 
  \left(
  \p^{\mu} \theta + \frac{NK^\prime_0}{2K_0} \delta^{\mu 0}
  \right)
  \phi^\+ T^a \phi 
  \right.
  \\
  \left.
  + i\alpha (1+\beta)
  \left(
    \phi^\+ T^a \p^\mu \phi -  \p^\mu\phi^\+ T^a \phi  +
  2 \phi^\+ T^a \phi  \p^\mu\phi^\+ \phi
  \right)
  \Big\}
  \right.
  \\
  \left.
  - \frac{4}{N} \alpha \gamma (1+\beta)
  \left(
   \p^{[\mu} \phi^\+ T^a \phi
  + \phi^\+ T^a \p^{[\mu} \phi
  \right)
  \left(
  \p^{\nu]} \theta + \frac{NK^\prime_0}{2K_0} \delta^{\nu] 0}
  \right)
  \right.
  \\
  \left.
  -  \frac{i}{2}\alpha^2 (1+\beta^2)
  \left(
   \phi^\+ T^a \phi \p^{[ \mu } \phi^\+  \p^{ \nu] }  \phi
   +  \p^{[ \mu } \phi^\+ T^a  \p^{ \nu] } \phi
   + \phi^\+ T^a \p^{[\mu} \phi \p^{\nu]}\phi^\+ \phi
   + \p^{[\mu} \phi^\+ T^a \phi \p^{\nu]}\phi^\+ \phi
  \right)
  \right].
  \label{eq:non-abelian-vorticity}
\end{multline}

Now we estimate the relative importance of terms in 
(\ref{eq:non-abelian-vorticity}).
Let us define a parameter $\epsilon \equiv D/C$.
Since $\epsilon$ expresses the deviation of the profile function from
the ground state value, $\epsilon$ is small away from the vortex core, as noted above.
We can estimate the strength of each term of the vorticity in terms of
the order of $\epsilon$: 
$\alpha(r) \sim \mathcal{O}(\epsilon^2)$, $\beta(r) \sim \mathcal{O}(\epsilon^0)$ and 
$\gamma(r) \sim \mathcal{O}(\epsilon)$. Therefore the leading order part of
the vorticity is given by
\begin{equation}
  \omega^a_{\lambda \sigma} 
  =  
  \epsilon_{\lambda \sigma \mu \nu}
  \p^\nu 
  \left[
  -\frac{16}{N} \gamma 
  \left\{ 
  \p^\mu \theta + \frac{NK^\prime_0}{2K_0} \delta^{\mu 0} 
  \right\}
  \phi^\+ T^a \phi
  \right] + {\cal O}(\epsilon^2).
\end{equation}

In summary, the interaction of vortices and quasiparticles, 
dual phonons and dual gluons, 
is described by the action 
\beq
  {\cal S}_{\rm int} = {\cal S}^{\rm Ph}_{\rm int}+{\cal S}^{\rm G}_{\rm int},
  \label{eq:interaction}
\eeq
respectively, 
in which ${\cal S}^{\rm Ph}_{\rm int}$ and ${\cal S}^{\rm G}_{\rm int}$ are defined by
\beq
  &&
  {\cal S}^{\rm Ph}_{\rm int}
  = -2\pi m^0 \int \diff^4 x~ B^0_{\mu\nu}
  \omega^{0,\mu\nu},
  \label{eq:action-phonon}
  \\
  &&
  {\cal S}^{\rm G}_{\rm int}
  =-\frac{m}{g} \int \diff^4 x~ B^a_{\mu\nu} \omega^{a,\mu\nu}.
  \label{eq:action-gluon}
\eeq
Here $\omega^0_{\mu\nu}$ and $\omega^a_{\mu\nu}$ are given by (\ref{eq:abelian-vorticity})
 and (\ref{eq:non-abelian-vorticity}), respectively.
Since these interaction terms do not couple to the space-time
metric, they are topological interactions.

The parameters $m$ and $m^0$ are chosen so that the kinetic term of
two-form fields are canonically normalized.
We require the kinetic terms of $B^a_{\mu\nu}$ and $B^0_{\mu\nu}$ to
take the form 
\begin{equation}
  \int \diff^4 x \left[
  - \frac{1}{12 \tilde{K}_{\mu\nu\sigma}} \left(
					 H^a_{\mu\nu\sigma}H^{a,\mu\nu\sigma}
  + 					 H^0_{\mu\nu\sigma}H^{0,\mu\nu\sigma}
  \right)
  \right],
\end{equation}
where 
$
H^a_{\mu\nu\sigma} \equiv 
\p_{\mu} B^a_{\nu\sigma} +  \p_{\nu} B^a_{\sigma\mu} +  \p_{\sigma} B^a_{\mu\nu} 
$,
$
H^0_{\mu\nu\sigma} \equiv 
\p_{\mu} B^0_{\nu\sigma} +  \p_{\nu} B^0_{\sigma\mu} +  \p_{\sigma} B^0_{\mu\nu} 
$
and
$
\tilde{K}_{\mu\nu\sigma} = \epsilon_{\rho\mu\nu\sigma} K^{\rho}
$.
Under this requirement, the parameters $m$ and $m^0$ are determined to be
\begin{equation}
  m = \frac{g\sqrt{C(r)}}{4}, \quad 
  m^0 = \frac{M}{\sqrt{2}} = \sqrt{ \frac{NC(r)+D(r)}{2}}.
\end{equation}

The interaction is localized around the vortex while gluons and phonons
propagate in the bulk.
This means that the vortex actually appears as a source for these particles.
Each term in (\ref{eq:non-abelian-vorticity}) is proportional to
$\alpha(r)$ or $\gamma(r)$ which are functions of the radial coordinate
and is nonzero only in the vicinity of the vortex. 
The functions $\alpha(r)$ or $\gamma(r)$ decay exponentially away from
the center of the vortex with the characteristic length, min$(m^{-1}_{\rm G},
m^{-1}_\chi)$, where $m_{\rm G}$ is the mass of gluons and $m_\chi$ is the mass
of traceless part of the scalar field $\Phi$ \cite{Eto:2009kg}.
Specific form of the profile functions $\alpha(r)$, $\beta(r)$ and $\gamma(r)$ can be
determined by solving the equations of motions numerically.

Now we discuss some properties of the interaction (\ref{eq:interaction}).
First let us look at the interaction (\ref{eq:action-phonon}) of vortices
with $\U(1)_\B$ phonons. 
This part is essentially the same as a vortex in a superfluid.
For general vortex configurations we can write the Abelian component of
the vorticity tensor as 
\beq
  (\omega^0)^{\mu\nu}(x) = 
  \frac{1}{N}\int \diff \tau \diff \sigma
  \frac{\p (X^\mu, X^\nu)}{\p (\tau,
  \sigma) } \delta^{(4)} (x - X^\mu(\tau, \sigma)),
\eeq
where the the space-time position of 
the vortex world sheet is parametrized as
$X^\mu(\tau, \sigma)$ with the world-sheet coordinates 
$\tau, \sigma$. 
The interaction of vortices with $\U(1)_\B$ phonons is
written as
\beq
  {\cal S}^{\rm Ph}_{\rm int} = -\frac{2\pi m^0}{N} 
  \int \diff \sigma^{\mu\nu} B^0_{\mu\nu}, \label{eq:phonon}
\eeq
where 
$
\diff \sigma^{\mu\nu} \equiv \frac{\p(X^\mu,X^\nu)}{\p(\tau, \sigma)}
\diff \tau \diff \sigma
$
 is an area element of the vortex world sheet. 
The magnitude of the Abelian vorticity is proportional to the winding number with
respect to $\U(1)_\B$, so we have the factor $1/N$ which is the winding
number of vortices of the lowest energy in the CFL phase.
$\U(1)_\B$ phonons $B_{\mu\nu}^0$ do not couple to the orientational zero modes on the
vortices, which is anticipated since $B_{\mu\nu}^0$ is a singlet under 
$\SU(N)_{\C+\F}$
while the orientational zero modes are in the fundamental representation.

Next, let us look at the interaction of vortices with 
gluons (\ref{eq:action-gluon}).
As can be seen in (\ref{eq:non-abelian-vorticity}), 
gluons actually couple to the orientational zero modes on the vortex.
As a result, virtual gluons can be emitted and absorbed from
orientational zero modes \footnote{
It is not possible to excite an on-shell gluon
in terms of the $\mathbb{C}P^{N-1}$ effective theory.
In order to describe the radiation of on-shell gluons from orientational
modes, we have to include terms with higer derivatives in the
$\mathbb{C}P^{N-1}$ effective theory.
However, we can still discuss the scattering of on-shell gluons by
vortices in terms of the Lagrangian obtained in the paper,
since there is no need to excite on-shell gluons in this case.
}.
This will lead to physical effects.
For example, when two vortices are in a small distance, the
orientational zero modes residing on them can exchange virtual gluons, 
which results in orientation-dependent corrections to the vortex-vortex interaction.
It is also possible that the orientational zero modes on a single vortex
emit and absorb virtual gluons. 
In the case of Abelian vortices, it is known that the emission and
absorption of virtual photons from a single vortex result in the
renormalization of its tension \cite{Orland:1994qt}. 
The emission and
absorption of virtual gluons can lead to similar effects for non-Abelian
vortices.

\section{Summary and Discussion}\label{sec:summary}
We have derived a dual Lagrangian of the Ginzburg-Landau 
effective Lagrangian for the CFL phase, in which 
massive gluon fields have been dualized 
to massive non-Abelian two-form fields while 
the $\U(1)_\B$ phonon field has been dualized 
to the massless Abelian two-form field.
In the dual theory, 
non-Abelian vortices have appeared as sources of these dual fields, and
we have thus obtained the topological interactions of non-Abelian vortices
with these quasiparticles. 
By making use of a single non-Abelian vortex solution, 
we have explicitly calculated 
the vorticity tensor of the single non-Abelian vortex, 
and then have found that the phonons couple to the translational zero modes
of the vortices as in the Abelian vortices 
while the gluons couple to the internal orientational zero modes ${\mathbb C}P^{N-1}$ of them.

Let us make comments on some applications of the dual Lagrangian
obtained above.

We can investigate the nature of the force between vortices. 
When vortices are separated at a large distance, $\U(1)_\B$ phonons mediate
the most part of the force between them;
the force between 
two vortices at the distance $R$ is proportional to $1/R$ 
\cite{Nakano:2008} which can be understood as follows. 
The tension of individual vortex with the $\U(1)_\B$ winding $k$ is 
$E \sim k^2 \log \Lambda$ with the system size $\Lambda$.
Therefore the tension of a non-Abelian vortex is $1/N^2$ 
of that of a $\U(1)_\B$ vortex. 
Consequently the interaction energy for 
two non-Abelian vortices 
and likewise the force between them 
is $1/N^2$ of those between two Abelian $\U(1)_\B$ vortices.
This is consistent with our result (\ref{eq:phonon}) 
of the coupling of $\U(1)_\B$ phonons 
to non-Abelian vortices, which is $1/N$ of the one 
of $\U(1)_\B$ phonons to Abelian vortices; 
the force between two non-Abelian vortices 
is proportional to the product of the $\U(1)_\B$ charges 
of them, which is $1/N^2$.

On the other hand, when the separation of vortices
is small,  exchange of virtual gluons will also contribue to the
force between vortices. 
In that case the property of the force is dependent on 
the ${\mathbb C}P^{N-1}$ internal orientations 
$\phi$ of the vortices 
because gluons (non-Abelian two-form fields) couple to 
the internal orientations through (\ref{eq:non-abelian-vorticity}).
This is natural since non-Abelian vortices carry color magnetic fluxes.

It is also interesting to apply the dual Lagrangian we have obtained to
the physics of the core of dense stars.
It is well known that the rotation of ultra cold atomic superfluids 
leads to a vortex lattice structure.
Neutron stars are also rotating rapidly, so we can expect that a 
lattice structure of non-Abelian vortices will be formed if the core is dense
enough to accommodate the CFL matter.
If the speed of rotation is high, the distance between neighboring vortices
is small and internal orientation will affect the nature of the
interaction of vortices.
As the type of collective structures of vortices is sensitive to details of
the interaction between them, it is interesting to investigate the phase
diagram as a function of the speed of rotations by making use of the action
we have derived. 
We expect that there may appear phases with vortex lattice structures 
different from ordinary triangular Abrikosov lattice.

The non-Abelian quantum turbulence phase where vortices are 
tangling and reconnecting here and there will be also possible if there is
some mechanism which leads to an instability of lattice structures \footnote{
It has been shown in \cite{Eto:2006db} that,
in the case of local non-Abelian vortices in 
the model with gauged $\U(1)_\B$ symmetry,
collisions of two vortices always result in their reconnection 
even when they have different internal orientations.
The same should hold for non-Abelian semisuperfluid vortices because 
whether the reconnection occurs or not depends on only topology.
}.
It is interesting to investigate the property of this turbulence since
we can expect behaviors which can be considerably different from
atomic superfluids or helium superfluids.
The energy spectra of turbulent fluids are determined by what kind of
interaction transfers the energy from lower to higher momentum region. 
In a helium superfluid the Kelvin wave cascade 
and the phonon emission are considered as the dominant
mechanism of energy transfer at the length scale which is smaller than
the mean vortex distance.
On the other hand, in the non-Abelian quantum turbulence,
not only the Kelvin wave cascade and the emission of
the $\U(1)_\B$ phonons from the vortex 
but also waves of the ${\mathbb C}P^{2}$ orientational zero modes 
can contribute to the energy cascade.
The propagation of $\mathbb{C}P^{2}$ waves can be affected by emission
and absorption of virtual gluons.
The dual Lagrangian obtained in this paper will provide a basis for the
analysis of properties of the non-Abelian quantum turbulence.

Finally the method we have used here can be applied to other types of
non-Abelian vortices. 
In the models with gauged $\U(1)$ symmetry, 
non-Abelian vortices also exist which are called local non-Abelian vortices. 
In the literature, local non-Abelian vortices in 
relativistic field theories with (or without) 
supersymmetry have been studied extensively \cite{local} 
(see \cite{local-review} for a review).
In this case, vortices are coupled to only massive fields \footnote{
When the number of flavors is greater than the number of color,
vortices are called semilocal.
In this case, semilocal non-Abelian vortices are also
coupled to massless Nambu-Goldstone bosons
corresponding to spontaneously broken flavor symmetry \cite{Shifman:2006kd} as well as massive particles.
}.
On the other hand,
at sufficiently high density where $\U(1)_{\rm A}$ symmetry is effectively
restored,
global non-Abelian vortices \cite{global} become topologically stable. 
They are expected to interact with CFL mesons topologically.
Via a dual transformation, 
CFL mesons are described by massless non-Abelian two-form fields 
with antisymmetric tensor gauge invariance,  
which is known as the Freedman-Townsend model \cite{Freedman:1980us}.
Unlike the case of semisuperfluid non-Abelian 
vortices,
interaction between two non-Abelian global vortices 
was shown to depend on their ${\mathbb C}P^{N-1}$ orientations
\cite{Nakano:2007dq}. 
This should be explained by the exchange of massless non-Abelian 
two-form fields.
Extensions to these cases will be reported elsewhere.

\begin{acknowledgments} 
M.~N. would like to thank Hiroaki Nakajima for a discussion at the early stage of this work, and to Minoru Eto for a useful comment. 
Y.~H. is grateful to T.~Hirano, T.~Hatsuda and S.~Sasaki for
 their encouragement and useful comments.
The work of M.~N. is partially supported by a Grant-in-Aid for Scientific Research No. 20740141 from 
the Ministry of Education, Culture, Sports, Science and Technology, Japan.
T.~K. is supported by the Japan Society for the Promotion of Science for Young Scientists.
\end{acknowledgments}

\appendix
\section{Derivation of  (\ref{eq:phi-j}) and (\ref{eq:j-phi})}

Here we derive Eqs.~(\ref{eq:phi-j}) and (\ref{eq:j-phi}).
We make use of the property of Casimir operators.
The generators of $\SU(N)$ in the fundamental representation obey the
following relation
\begin{equation}
  (T^a)^i_j(T^a)^k_l = \frac{1}{2} \delta^i_l \delta^k_j - \frac{1}{2N}
  \delta^i_j \delta^k_l.
  \label{eq:casimir2}
\end{equation}

First we show that $(\bm{\Phi}^{-1})^{ab}_{\mu\nu}$ can be written as
\begin{equation}
  (\bm{\Phi}^{-1})^{ab}_{\mu\nu} = \frac{4g_{\mu\nu}}{C\sqrt{K_\mu K_\nu}}
  \left\{
  2 \Tr (T^aT^b) - 2(1-\beta) \phi^\+ T^aT^b \phi + 2(1-\beta-\gamma)
  \phi^\+ T^a \phi \phi^\+ T^b \phi 
  \right\},
  \label{eq:phi-inverse}
\end{equation}
with $\beta(r) \equiv \frac{C}{C+D}$ and  $\gamma(r) \equiv
\frac{D}{C+D(1-1/N)}$.
Let us define the quantity
\begin{equation}
 \begin{split}
  X^{ab}_n \equiv \phi^\+ T^a T^{a_1} \phi \phi^\+ T^{a_1} T^{a_2} \phi \cdots 
  \phi^\+ T^{a_{n-1}} T^{b} \phi.
 \end{split} 
\end{equation}
By using the property of the Casimir operator (\ref{eq:casimir2}) and
$\phi^\+\phi=1$ it can be shown that $X_n$ satisfy the following recurrence
relation
\begin{equation}
  X^{ab}_n = \frac{1}{2} X^{ab}_{n-1} 
  - 
  \frac{1}{2N} \left( \frac{1}{2} - \frac{1}{2N} \right)^{n-2} \phi^\+ T^a
  \phi \phi^\+ T^b \phi.
\end{equation}
This equation can be solved to give
\begin{equation}
  X^{ab}_n = \left( \frac{1}{2} \right)^{n-1} 
  \left[
  \phi^\+ T^a T^b \phi - 2 \left\{ 1 - (1-1/N)^{n-1} \right\}  \phi^\+ T^a
  \phi \phi^\+ T^b \phi
  \right],
\end{equation}
with the initial condition $X^{ab}_1 = \phi^\+ T^a T^b \phi$.
The quantity $(\bm{\Phi}^{-1})^{ab}_{\mu\nu}$ is defined by an expansion
\begin{equation}
 \begin{split}
  (\bm{\Phi}^{-1})^{ab}_{\mu\nu} &= g_{\mu\nu}\frac{1}{\sqrt{K_\mu K_\nu}} \frac{4}{C}
  \left(\delta^{ab} + 
  \sum_{n=1}^{\infty} \left(-\frac{2D}{C}\right)^n
  X^{ab}_n
  \right).
  \label{eq:XofN}
 \end{split}
\end{equation}
Substituting Eq.~(\ref{eq:XofN}) into the above equation, now we perform
the sum with respect to $n$ and obtain the result
(\ref{eq:phi-inverse}).

Then let us proceed to calculate $\left( \bm{\Phi}^{-1}J\right)^a_\mu =
\left( \bm{\Phi}^{-1}\right)^{ab}_{\mu\nu}J^{b,\nu} $.
The current $J^a_\mu$ for $a = 1, \cdots, N^2-1$ can be written in terms
of ${\cal F}$ as Eq.~(\ref{eq:current-with-f}).
Note that the following relations hold
\beq
 \phi^\+ T^a T^b \phi \phi^\+ T^b \phi = \frac{1}{2}\left(1 - \frac{1}{N}\right) \phi^\+ T^a \phi,
\eeq
\begin{equation}
 \begin{split}
  \phi^\+ T^aT^b \phi \Tr \left\{ T^b {\cal F}_\mu(-1,1)\right\} 
  &=\frac{1}{2} \phi^\+ T^a  {\cal F}_\mu(-1,1) \phi - \frac{1}{2N}
  \phi^\+ T^a \phi \Tr {\cal F}_\mu(-1,1) 
  \\
  &=\frac{1}{2} \Tr \left\{ T^a  {\cal F}_\mu(0,1) \right\},
 \end{split}
\end{equation}
\begin{equation}
 \begin{split}
  \phi^\+ T^b \phi \Tr \left\{ T^b {\cal F}_\mu(-1,1)\right\} 
  &=\frac{1}{2} \phi^\+   {\cal F}_\mu(-1,1) \phi - \frac{1}{2N} \Tr {\cal F}_\mu(-1,1) 
  \\
  &=0,
 \end{split}
\end{equation}
where we have used Eq.~(\ref{eq:casimir2}), $\phi^\+\phi=1$ and the
properties of ${\cal F}$. 
Applying these relations to the product of Eqs.~(\ref{eq:phi-inverse}) and
(\ref{eq:current-with-f}) leads to
\beq
  (\bm{\Phi}^{-1}J)^a_\mu = -\frac{8}{N} \gamma \phi^\+T^a\phi 
  \left\{ 
  \p_{\mu} \theta + \frac{NK^\prime_0}{2K_0} \delta_{\mu 0}
  \right\}
  + i \alpha \Tr \left[ T^a {\cal F}_\mu( -1, \beta ) \right],
\eeq
with $\alpha \equiv \frac{4G^2}{C}$.
By taking the complex conjugate,  we have
\beq
  (J\bm{\Phi}^{-1})^a_\mu = -\frac{8}{N} \gamma \phi^\+ T^a \phi 
  \left\{ 
  \p_{\mu} \theta + \frac{NK^\prime_0}{2K_0} \delta_{\mu 0}
  \right\}
  - i \alpha \Tr \left[ T^a {\cal F}_\mu( \beta,- 1 ) \right].
\eeq

\end{document}